\documentclass[twocolumn]{aastex631}

\hypersetup{linkcolor=red, citecolor=blue, filecolor=cyan, urlcolor=magenta}
\usepackage{multirow}
\usepackage{amsmath}

\accepted{September 21, 2023}

\submitjournal{ApJ}

\shorttitle{COSMOS-ReGEM: I. Merging Galaxies}
\shortauthors{J. Ren}
\graphicspath{{./}{figures/}}
\begin{document}

\email{E-mail: nan.li@nao.cas.cn, fsliu@nao.cas.cn}

 \title{Revisiting Galaxy Evolution in Morphology in the COSMOS field (COSMOS-ReGEM): \\
 I. Merging Galaxies}

\author{Jian Ren}
\affil{National Astronomical Observatories, Chinese Academy of Sciences, 20A Datun Road, Chaoyang District, Beijing 100101, China}
\affil{Key Laboratory of Space Astronomy and Technology, National Astronomical Observatories, Chinese Academy of Sciences, 20A Datun Road, Chaoyang District, Beijing 100101, China} 

\author{Nan Li $^{\color{blue} \dagger}$}
\affil{National Astronomical Observatories, Chinese Academy of Sciences, 20A Datun Road, Chaoyang District, Beijing 100101, China}

\affil{Key Laboratory of Space Astronomy and Technology, National Astronomical Observatories, Chinese Academy of Sciences, 20A Datun Road, Chaoyang District, Beijing 100101, China}
\affil{School of Astronomy and Space Science, University of Chinese Academy of Science, Beĳing 100049, China}

\author{F. S. Liu $^{\color{blue} \dagger}$}
\affil{National Astronomical Observatories, Chinese Academy of Sciences, 20A Datun Road, Chaoyang District, Beijing 100101, China}
\affil{Key Laboratory of Optical Astronomy, National Astronomical Observatories, Chinese Academy of Sciences, 20A Datun Road, Chaoyang District, Beijing 100101, China}
\affil{School of Astronomy and Space Science, University of Chinese Academy of Science, Beĳing 100049, China}

\author{Qifan Cui}
\affil{National Astronomical Observatories, Chinese Academy of Sciences, 20A Datun Road, Chaoyang District, Beijing 100101, China}
\affil{Key Laboratory of Space Astronomy and Technology, National Astronomical Observatories, Chinese Academy of Sciences, 20A Datun Road, Chaoyang District, Beijing 100101, China}

\author{Mingxiang Fu}
\affil{School of Astronomy and Space Science, University of Chinese Academy of Science, Beĳing 100049, China}
\affil{Key Laboratory of Space Astronomy and Technology, National Astronomical Observatories, Chinese Academy of Sciences, 20A Datun Road, Chaoyang District, Beijing 100101, China}

\author{Xian Zhong Zheng}
\affil{Purple Mountain Observatory, Chinese Academy of Sciences, 10 Yuanhua Road, Nanjing 210034, China}
\affil{School of Astronomy and Space Sciences, University of Science and Technology of China, Hefei 230026, China}

\begin{abstract}

We revisit the evolution of galaxy morphology in the COSMOS field over the redshift range $0.2\leq z \leq 1$, using a large and complete sample of 33,605 galaxies with a stellar mass of log($M_{\ast}$/M$_{\odot} )>9.5$ with significantly improved redshifts and comprehensive non-parametric morphological parameters. Our sample has 13,881 ($\sim41.3\%$) galaxies with reliable spectroscopic redshifts and has more accurate photometric redshifts with a $\sigma_{\rm NMAD} \sim 0.005$. This paper is the first in a series that investigates merging galaxies and their properties. We identify 3,594 major merging galaxies through visual inspection and find 1,737 massive galaxy pairs with log($M_\ast$/M$_\odot$)$>10.1$. Among the family of non-parametric morphological parameters including $C$, $A$, $S$, $Gini$, $M_{\rm 20}$, $A_{\rm O}$, and $D_{\rm O}$, we find that the outer asymmetry parameter $A_{\rm O}$ and the second-order momentum parameter $M_{\rm 20}$ are the best tracers of merging features than other combinations. Hence, we propose a criterion for selecting candidates of violently star-forming mergers:  $M_{\rm 20}> -3A_{\rm O}+3$ at $0.2<z<0.6$ and $M_{\rm 20}> -6A_{\rm O}+3.7$ at $0.6<z<1.0$. Furthermore, we show that both the visual merger sample and the pair sample exhibit a similar evolution in the merger rate at $z<1$, with $\Re \sim(1+z)^{1.79 \pm 0.13}$ for the visual merger sample and $\Re \sim(1+z)^{2.02\pm 0.42}$ for the pair sample. The visual merger sample has a specific star formation rate that is about 0.16\,dex higher than that of non-merger galaxies, whereas no significant star formation excess is observed in the pair sample. This suggests that the effects of mergers on star formation differ at different merger stages.

\end{abstract}


\section{Introduction} \label{sec:intro}

Galaxy mergers are one of the main mechanisms of stellar mass growth in galaxies. Since the observation of Arp galaxies by \citet{Arp1966} and their initial reproduction in numerical simulations by \citet{Toomre1972}, galaxy mergers have been widely studied for the last half-century. Current research focuses on the effects of mergers on galaxy properties, including star formation \citep[e.g.][]{Mihos1996, Teyssier2010, Scudder2012, Pearson2019, Ellison2022}, AGN activity \citep{Ellison2011, Treister2012, Satyapal2014, Barrows2017, Ricci2017, Byrne-Mamahit2023}, size growth \citep{Naab2009, Nipoti2009, Bluck2012, Newman2012, Sanjuan2012}, morphological transformation \citep{Naab2006, Bournaud2011, Rodriguez-Gomez2017, Martin2018}, and dynamic features \citep{Martel2020, Li2021}. Additionally, accurate measurement of the merger rate is the best way to understand the impact of mergers on galaxy formation and evolution on a cosmological timescale \citep{Lacey1993, Conselice2008, Shi2009}.

The first step in studying merging galaxies is to obtain a complete sample of mergers. Currently, the most commonly used merger identification methods include visual identification, galaxy pairs, Machine Learning (ML), and non-parametric morphological methods. The visual identification method selects merging galaxies 
based on their merging features, such as tidal tails, multiple nuclei, and disturbed morphologies. In recent years, the ``Galaxy Zoo" project \citep{Lintott2008} and other works \citep{Jogee2009, Bridge2010, Kartaltepe2015, Simmons2017, Mata2022, Ryan2023} have published visual merger samples. The galaxy pair method usually selects close pairs whose relative velocities and nuclear separations are below a certain threshold \citep[e.g.][]{Patton2000, Lambas2003, Lin2004, Kartaltepe2007, Ellison2008, Xu2012, Mantha2018, Duncan2019}. This method generally requires spectroscopic redshifts or high-precision photometric redshifts to reduce the influence of projection effects \citep{deRavel2009, Lopez-Sanjuan2015, Rodriguez2020}. The ML methods have emerged in the last decade and have become increasingly dominant for automatic galaxy morphological classification and merging galaxy identification \citep[e.g.][]{Dieleman2015, Huertas-Company2015, Ackermann2018, Vavilova2021}. Supervised learning algorithms require additional methods to obtain labeled training samples with higher completeness, while unsupervised learning algorithms have also been rapidly developed \citep{Schutter2015, Hocking2018, Hausen2020, Martin2020, Cheng2021}.

Non-parametric galaxy morphological methods do not rely on a specific analytical function of the galaxy's light distribution and can therefore be applied to the classification of irregular and standard Hubble-type galaxies. \citet{Abraham1994, Abraham1996} introduced the light concentration parameter $C$ and \citet{Schade1995} proposed the rotational asymmetry $A$ as a way to automatically distinguish between early- and late-type galaxies and to classify irregular and merging galaxies. Later, $C$ and $A$ were improved for better surface brightness selection and determination of galaxy centers by \citet{Bershady2000} and \citet{Conselice2000}. The galaxy smoothness parameter $S$ was proposed by \citet{Takamiya1999} and \citet{Conselice2003a} to study galaxy colors and star formation activity. These three parameters together formed the first non-parametric morphological method, $CAS$. \citet{Lotz2004} introduced the $Gini$ coefficient ($G$) into galaxy morphological studies and invented the $Gini-M_{\rm 20}$ method, which is useful for selecting merger candidates. \citet{Wen2014} found that galaxy bulges have a significant influence on asymmetry $A$ measurement and thus calculated the outer asymmetry $A_{\rm O}$ after subtracting 50\% of the total light from the galaxy center. They also computed the deviation in the center of the inner and outer half-light region, obtaining $D_{\rm O}$. Then, they introduced a novel non-parametric method, $A_{\rm O}-D_{\rm O}$, which helps identify merging galaxies with long tidal tails \citep{Wen2016}. The accuracy and completeness of mergers identified by current non-parametric methods are relatively low. This mainly stems from the fact that different non-parametric methods focus on different merger features. To improve the accuracy of non-parametric-based merger identification methods, a large and complete merger sample is needed.

In recent years, many high-precision spectroscopic and photometric data have been available in the COSMOS field, which greatly improving our understanding of galaxy formation and evolution at intermediate to high redshifts. Hence, we construct a large and complete sample of galaxies to revisit the morphological evolution of galaxies over $0.2 \le z \le 1$, with improved redshifts and comprehensive non-parametric morphological parameters measured by ourselves. This paper is the first in a series that investigates merging galaxies and their properties. 

The outline of this paper is as follows. In Section 2, we describe our data and sample selection. In Section 3, we describe the methods used to identify the visual merger sample and the pair sample, and the measurement of non-parametric morphological parameters. In Section 4, we present our main results. We give some discussions in Section 5 and finish with a summary and conclusion in Section 6. Throughout this paper, we utilize a concordance cosmology with $H_{0}=$70 km s$^{-1}$ Mpc$^{-1}$, $\Omega_{\rm m}$=0.3, and $\Omega_{\Lambda}$=0.7. All photometric magnitudes are given in the AB system.

\begin{figure*}[ht!]
\centering
\includegraphics[width=1\textwidth] {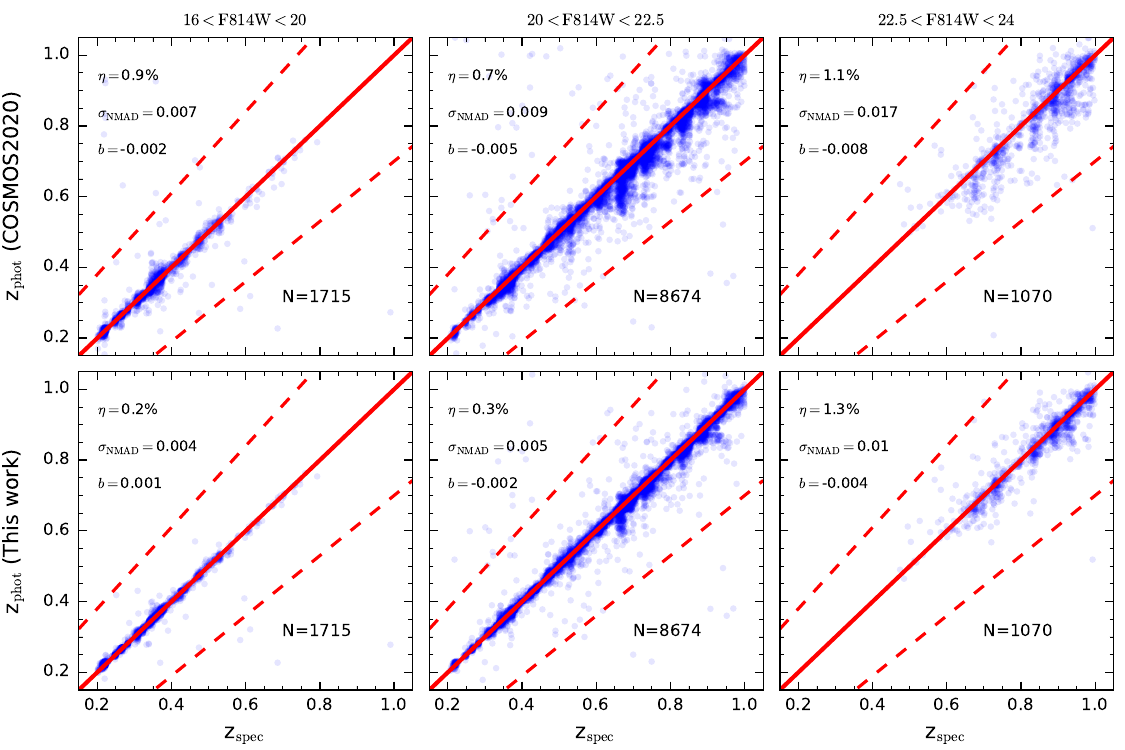}
\caption{ The top three panels show the comparison between spec-$z$  and the most reliable phot-$z$ in the COSMOS2020/CLASSIC sample, while the bottom three panels show the phot-$z$ in our work. Galaxies with $|z_{\rm phot}-z_{\rm spec}|>0.15(1+z_{\rm spec})$ are considered as outliers \citep{Hildebrandt2012}. Phot-$z$ error is defined by \citet{Hoaglin1983} and denoted by $\sigma_{\rm NMAD}$.}
\label{fig:fig001}
\end{figure*}   

\section{Data} \label{sec:data}

\subsection{Photometry Catalogs and Imaging Data} \label{sec:data}

The Cosmic Evolution Survey \citep[COSMOS;][]{Scoville2007} aims to study galaxy evolution, star formation activity, AGN, cosmic large-scale structure, and dark matter at redshifts of $0.5<z<6$. It covers an area of 2 deg$^2$. In recent years, ground-based and space-based telescope observations have obtained a large amount of multi-wavelength data from X-ray to radio bands.

Since the COSMOS2015 photometry redshift catalog \citep{Laigle2016} was publicly available, a wealth of new photometric and spectroscopic observations have been conducted in the COSMOS field. \citet{Weaver2022} collected available multi-wavelength imaging data from the COSMOS field before 2020. They obtained the COSMOS2020/CLASSIC and COSMOS2020/FARMER catalogs by using the traditional aperture photometric method and FARMER profile-fitting photometric extraction method to detect sources, respectively. The COSMOS2020/CLASSIC catalog contains 1.7 million sources across the 2 deg$^2$ area. Photometric redshifts and galaxy physical parameters were obtained using the \texttt{Le Phare} \citep{Arnouts1999, Arnouts2002, Ilbert2006} and \texttt{EAZY} \citep{Brammer2008} SED fitting codes. For galaxies with $i<22.5$ mag, the phot-$z$ accuracy achieves $\sigma \sim 0.01(1+z)$. The stellar mass limit (M$_{\rm lim}$) of the complete sample is 0.5\,dex better than previous catalogs.

The {\it HST}/ACS survey covers an area of 1.64 deg$^2$ and provides high-resolution F814W imaging data \citep{Koekemoer2007}, which has greatly promoted the study of rest-frame optical band morphology and structure evolution for galaxies with $z\leq1$. The pixel scale of the imaging data and FWHM of the PSF are $0.^{\prime\prime}03$ per pixel and $0.^{\prime\prime}09$, respectively. The imaging data reach a limiting magnitude of 25.6\,mag for extended sources within a circular aperture radius of $0.^{\prime\prime}3$ (10 pixels).

\subsection{Redshift data}
We recompile the catalog of spectroscopic redshifts in the COSMOS fields. There are several large spectroscopic surveys publicly available. The first is the zCOSMOS survey \citep{Lilly2007, Lilly2009}, which allocated 600 hr of observation and is divided into a bright and a deep component.

The zCOSMOS-bright is a magnitude-limited I-band $I_{\rm AB} < 22.5$\,mag sample of about 20,000 galaxies with $0.1 < z < 1.2$ covering the entire 1.7 deg$^2$ COSMOS ACS field. A part of the VIMOS VLT Deep Survey \citep[VVDS,][]{LeFevre2013} includes the COSMOS field and it contains about 35,000 samples of I-band magnitude reaching   $I_{\rm AB}=24.75$\,mag. The VIMOS Ultra Deep Survey \citep[VUDS,][]{LeFevre2015, Tasca2017} is a spectroscopic redshift survey of $\sim$10,000 very faint galaxies to study the main phase of galaxy assembly at $2 < z < 6$ covering some COSMOS regions. The ESO-VLT also has several other surveys. \citet{Rosani2020} used a Multi Unit Spectroscopic Explorer (MUSE) to study faint star-forming galaxies at $z < 1.5$ and Ly$\alpha$ emitters at $z > 3$. The Large Early Galaxy Astrophysics Census (LEGA-C) is targeting $0.6 < z < 1.0$, Ks-selected galaxies with spectra measuring stellar velocity dispersions and stellar population properties \citep[e.g.][]{Straatman2018, vanderWel2021}. The PRIsm MUlti-object Survey \citep[PRIMUS,][]{Coil2011} is a spectroscopic faint galaxy redshift survey up to $z\sim$1 using the Magellan I Baade telescope. The MOSFIRE Deep Evolution Field \citep[MOSDEF,][]{Kriek2015} survey aims to obtain $\sim$1500 galaxies at $1.37 < z < 3.80$ in three well-studied CANDELS fields. The Deep Imaging Multi-Object Spectrograph (DEIMOS) on the Keck II telescope selected a large sample of broad redshift distribution of about 10,000 objects at $0 < z < 6$ in the COSMOS field \citep{Hasinger2018}. The hCOSMOS redshift survey \citep{Damjanov2018} was conducted with the Hectospec spectrograph on the MMT, which measures redshifts and the age-sensitive Dn4000 index over the entire redshift interval $0.001 < z < 0.6$. The Complete Calibration of the Color-Redshift Relation (C3R2) survey is a multi-institution, multi-instrument survey that aims to map the empirical relation of galaxy color to redshift up to $I_{\rm AB} \sim$24.5\,mag. The C3R2 has released several data \citep[e.g.][]{Masters2017, Masters2019, EuclidCollaboration2020, Stanford2021} and about 10,000 highly-reliable redshifts. 
In addition, we also include the unpublished (e.g., UCR DEIMOS Survey) spectroscopic redshifts compiled by N. P. Hathi (2018, private communication).

We then cross-matched the COSMOS2020/CLASSIC photometry catalog with our spec-$z$ catalog using an aperture of $r= 0.54^{\prime\prime}$. Finally, we obtained spec-$z$ for 29562 galaxies, of which about 24,682 galaxies secure spectroscopic (flagged as ``very secure'' or ``reliable'') redshifts.

There is a large amount of phot-$z$ data available for the COSMOS field, which can help to improve the accuracy of phot-$z$ measurements. We collected the COSMOS Photometric Redshifts with 30-Bands \citep[CPR30,][]{Ilbert2009}, UrtaVISTA \citep{Muzzin2013}, and COSMOS2015 \citep{Laigle2016} phot-$z$ catalogs and matched them to the COSMOS2020/CLASSIC catalog. For the matched COSMOS2020/CLASSIC catalog, each source has between 1 and 5 phot-$z$ measurements. If a source has three or five phot-$z$, we take the median phot-$z$ as our phot-$z$. If there are four measurements, we remove an outlier and calculate the median phot-$z$. If a source has fewer than three phot-$z$, we use the \texttt{Le Phare} phot-$z$ in the COSMOS2020/CLASSIC catalog. 
Most of the bright sources ($I<24$\,mag) have at least four phot-$z$, so this approach significantly improves the phot-$z$ accuracy for bright sources.

Redshifts ($z_{\rm best}$) used in this work are in the order of priority of secure spec-$z$, the spec-$z$ with a deviation less than 3\,$\sigma$ from our phot-$z$, and the phot-$z$. Compared to the secure spec-$z$, the accuracy of our phot-$z$ is as high as $\sigma_{\rm NMAD} \sim 0.005$ for galaxies with $I_{\rm F814W }<22.5$ at spec-$z<1$, as shown in Figure \ref{fig:fig001}.

\begin{figure}[h!]
\centering
\includegraphics[width=1\columnwidth] {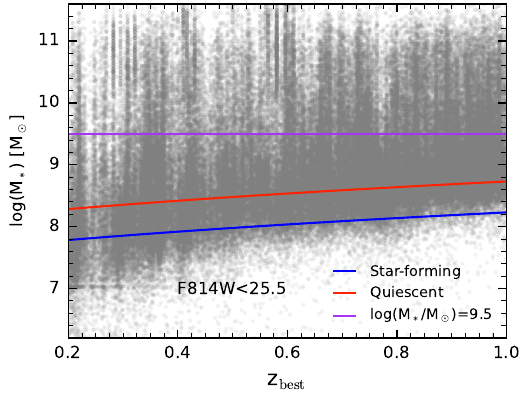}
\caption{Distribution of stellar masses and redshifts of the galaxies with $i<25.5$ at $0.2\leq z_{\rm best}\leq 1$  in the COSMOS {\it HST}/ACS F814W imaging survey covered area. The blue and red lines are the empirical stellar mass completeness for the star-forming and quiescent galaxy sample computed by \citet{Weaver2022}. Our sample is the galaxies above the magenta dotted line.
}
\label{fig:fig002}
\end{figure}

\subsection{Stellar mass}

The stellar mass determined by applying different SFHs in the SED fitting process may differ slightly. In this work, we first obtained two sets of stellar masses using the exponentially declining and delayed exponentially declining SFHs, respectively. We then combined these two sets of results with the UltraVISTA catalog to compute the median stellar masses. All three sets of stellar masses were fitted using FAST code \citep{Kriek2009} and based on a grid of Bruzual \& Charlot models \citep{Bruzual2003} that assume a Chabrier Initial Mass Function \citep[IMF,][]{Chabrier2003}, solar metallicity, and a Calzetti dust law \citep{Calzetti1994}. This method can effectively improve the accuracy and reduce the outliers of stellar masses \citep{Santini2015}. We did not use the stellar mass estimates in the COSMOS2015 and COSMOS2020 catalogs because they use complex SFHs in their library, resulting in stellar mass estimates that are $0.1–0.3$\,dex higher \citep{Leja2019, Weaver2022}.

\begin{figure*}[t]
\centering
\includegraphics[width=1\textwidth] {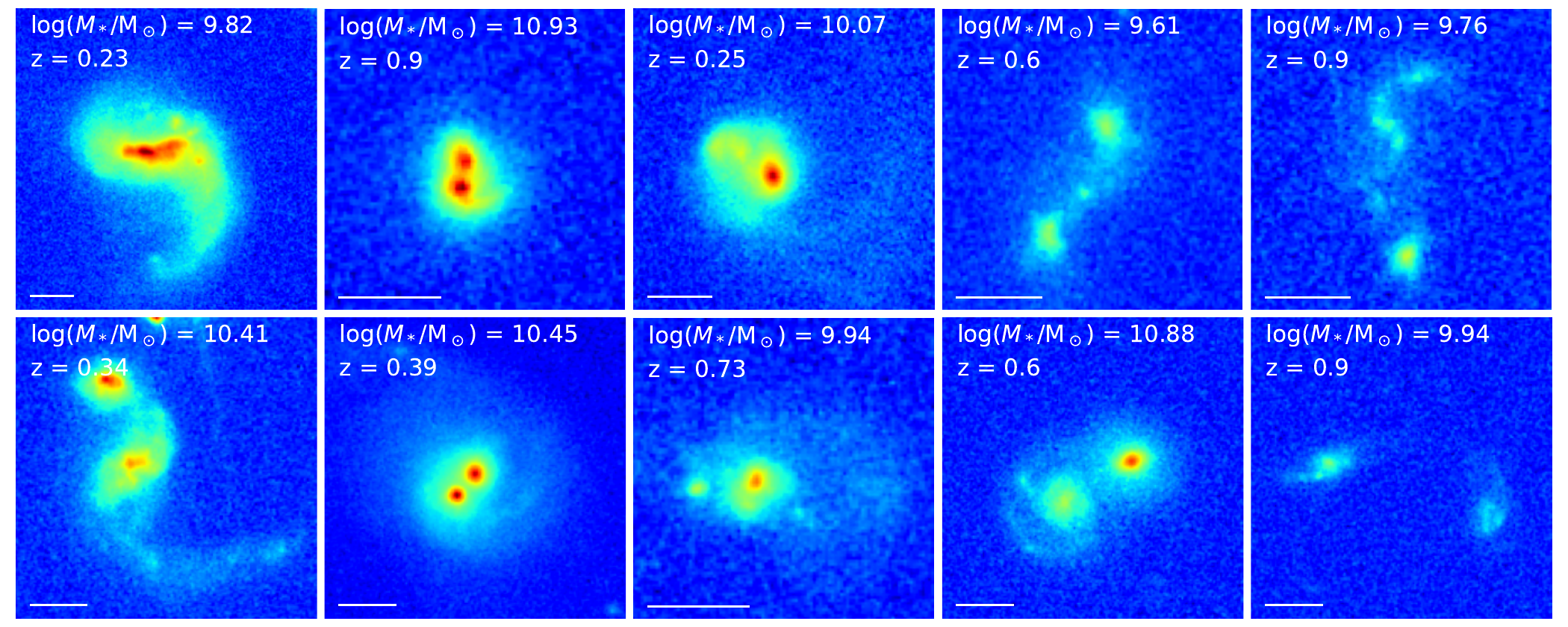}
\caption{Example images of merging galaxies. From the left two panels to the right two panels show the merging galaxies with tidal tails, double nuclei, merger remnants, close pairs, and pairs with disturbed morphologies respectively. The redshift and stellar mass are labeled on each image. The length of the white line in each image is $1^{\prime\prime}$. }
\label{fig:fig003}
\end{figure*}         

\subsection{Sample selection} \label{sec:data}

Our objective is to investigate the morphologies of merging galaxies using non-parametric methods and their properties. A complete galaxy sample with high-resolution {\it HST}/ACS images is necessary for this study. Therefore, we selected a volume-limited parent sample with log($M_\ast$/M$_\odot)\geq $9.5, and 0.2$\leq z_{\rm best}\leq $1.0 within the HST/ACS survey coverage area. Our sample comprises 33605 galaxies, of which about 41.3\% have spec-$z$. We matched the existing Av data to the parent sample. For each galaxy, we adopted the median Av as the best Av. We then used the A$_{\rm2800}$ \citep[A$_{\rm2800}$=1.8\,Av, ][]{Calzetti1994} and rest-frame L$_{\rm 2800}$ to calculate the star formation rates (SFRs) of the parent sample, SFR$_{\rm UV, cor}$=2.59$\times $10$^{-10}$L$_{\rm UV, cor}$[L$_\odot$] \citep{Kennicutt2012}. Figure \ref{fig:fig001} presents a comparison of the accuracy of phot-$z$ between our sample and the COSMOS2020/CLASSIC sample. Figure \ref{fig:fig002} illustrates our sample selection.


\section{Methodology} \label{sec:method}

\subsection{Visual Identifications of Merging Galaxies} 
\label{sec:visual}

We applied the visual identification method to search for mergers from the parent sample. Galaxies undergoing a merger often exhibit clear tidal features that distinguish them from normal Hubble-type galaxies. We identified mergers based on the following four features.

{\bf Tails and bridges: } Tidal tails and bridges are the most common merger features. Tidal tails form after the first encounter in mergers and remain observable until the merger is complete \citep[e.g.][]{Toomre1972, Barnes1992a, Ren2020}. Tidal bridges are often seen in close galaxy pairs. They are lower surface brightness stellar belts that connect two galaxies.

{\bf Double nuclei:} Double or multiple nucleus structures are frequently found during late-stage mergers when two galaxies have not yet merged their nuclei. This feature lasts for a shorter period compared to tidal tails. When searching for mergers with double nuclei, one should consider the overall galaxy morphology to avoid misidentifying SFGs with massive star-forming clumps as mergers.

{\bf Merger remnants:} Disturbed morphologies, tidal streams, shells, and other structures can also serve as merger tracers when tidal tails and double nuclei are not easily observed.

{\bf Visual pairs: } Pairs of galaxies already in contact or displaying significant morphological disturbances can also be identified as mergers.

The above morphologies represent the early to late stages of mergers. We have identified all major mergers with tidal features from the parent sample to the best of our ability. It is important to note that separated galaxy pairs without significant morphological disturbance will not be considered as merger systems. These pairs will be included in the pair sample only if they meet the close pair criteria listed in Section \ref{pairs}.

The visual identification of mergers was conducted independently by Ren, Fu, and Liu using F814W 501$\times$501\, pixel cutout images. Each galaxy was given a flag of 2, 1, or 0, indicating merger, suspected merger, and non-merger, respectively. A galaxy was considered a merger if the sum of flags from different individuals was greater than three. During the identification process, we also encountered a small number of minor mergers. In reality, minor mergers are more common; however, the merger processes associated with them rarely produce prominent morphological perturbations. Therefore, visual identification is not a reliable method for selecting minor mergers. Instead, researchers have used galaxy-pair methods to identify minor mergers \citep{Conselice2022}. For spatially resolved galaxy pairs, we removed the merger systems with $\Delta$log(M$_*$/M$_\odot)>0.6$. In total, we obtained a sample of 3594 major mergers.

\subsection{Massive Galaxy-Galaxy Pairs}
\label{pairs}

Given the abundance of spectral redshifts and the improved accuracy of photometric redshifts in our sample, we utilized the following method to identify close galaxy pairs.

{\bf Mass ratio:} We use $1<M_1/M_2<4$ to select major galaxy pairs. This mass ratio corresponds to $|\Delta$log($M_*$/M$_{\odot})|<0.6$. Since the lower mass limit of our sample is log($M_*$/M$_\odot)=9.5$, the stellar mass of the primary galaxy in pairs larger than 10$^{10.1}$M$_\odot$ is mass-complete. Notably, Brightest Cluster Galaxies (BCGs) are often associated with numerous less massive satellite galaxies, whose properties are substantially influenced by the central galaxy. Such small-mass satellite pairs may therefore affect the analysis of the merger rate and star formation properties.  

{\bf Projected separation:} Although different studies use slightly variable values of the projected separation (R$_{\rm proj}$) to identify galaxy pairs, any discrepancies in pair selection based on R$_{\rm proj}$ have minimal impacts on the calculated merger rate. Therefore, we adopted a criterion of $5<$ R$_{\rm proj}<50$\,kpc to select close galaxy pairs \citep{Mantha2018}.

{\bf Relative velocities:} Since our sample comprises both spectroscopic and photometric redshift galaxies, we employed different selection criteria for spectroscopic and photometric galaxy pairs. Specifically, we used a threshold of $|\Delta V|<500$ km\,s$^{-1}$ for spec-$z$+spec-$z$ pairs \citep{Kartaltepe2007, Patton2008, Lin2008}, and $|\Delta z|< \sigma_{z}$ for spec-$z$+phot-$z$ pairs, where $\sigma_z$ is the phot-$z$ error at fixed stellar mass and redshift bins, which is different from the $\sigma_{\rm NMAD}$ in Figure \ref{fig:fig001}. For phot-$z$+phot-$z$ pairs, we utilized the selection criterion of $|\Delta z|<\sqrt{\sigma_{z1}^2+\sigma_{z2}^2}$, which was introduced by \citet{Bundy2009} and frequently used in the selection of galaxy pairs in photometric redshift samples.

In total, we selected 1737 massive galaxy pairs, and the results are presented in Table \ref{tab: pair samples}. It is noted that some close galaxy pairs with tidal features exist in both the visual merger sample and the pair sample.

\begin{table}[]
\centering
\caption{The information of massive galaxy pair sample. }
\label{tab: pair samples}
 \begin{tabular}{|p{3cm}<{\centering}|p{3cm}<{\centering}|p{1.3cm}<{\centering}|}
\hline
Pairs type &Relative velocities & Number \\
\hline
spec-$z$+spec-$z$   & $|\Delta V|<500$ km\,s$^{-1}$ & 338 \\
\hline
spec-$z$+phot-$z$   & $|\Delta z|< \sigma_z$ & 768  \\
\hline
phot-$z$+phot-$z$ & $|\Delta z|<\sqrt{\sigma_{z1}^2+\sigma_{z2}^2}$ & 631 \\
\hline
 \end{tabular}
 \end{table}

\subsection{Shape Measurements of Parent Sample} \label{sec:Non-para}

We measure the non-parametric morphological parameters ($C$, $A$, $S$, $G$, $M_{\rm 20}$, $A_{\rm O}$, $D_{\rm O}$, and $R_{\rm 50}$) of the parent sample using a Python code developed by Ren et al. (in preparation), which will be released as a part of the China Space Station Telescope (CSST) data reduction pipeline.

To obtain these parameters, we need some auxiliary parameters. We first obtain the Petrosian radius \citep[R$_{\rm p}$,][]{Petrosian1976} of each galaxy, which is defined as the radius at which the mean surface brightness is equal to some fraction $\eta$ of the mean surface brightness within R$_{\rm p}$. For our measurements, we set $\eta$=0.2 to calculate the major axis of elliptical apertures as R$_{\rm p}$. In the non-parametric parameter measurements, we defined the total flux of a galaxy as the flux within 1.5\,R$_{\rm p}$ \citep{Lotz2004, Rodriguez-Gomez2019}. \citet{Lotz2004} points out that the $Gini$ coefficient is sensitive to the signal-to-noise ratio (SNR). Therefore, we used the method given by \citet{Lotz2004} to produce a $Gini\mbox{-}segmentation$ map. We first smooth the galaxy image using a Gaussian kernel with $\sigma=0.2$\,R$_{\rm p}$, and then we set the value of the pixel with a flux above the mean flux at R$_{\rm p}$ and below 10 $\sigma$ with neighboring pixels to 1, and other pixels to 0. The $Gini\mbox{-}segmentation$ map was used to measure $G$, $M_{\rm 20}$, and SNR.

\subsubsection{$CAS $ parameters}

$A$: The asymmetry parameter is expressed as the difference between the rotated image and the original image of galaxies. It is obtained by subtracting the $180^\circ$-rotated  image from the original image \citep{Conselice2000, Conselice2003a}, defined as, 

\begin{equation}
A={\rm min}( \frac{\sum |I_{0}-I_{180}|}{\sum|I_{0}|})-{\rm min}(\frac{\sum |B_{0}-B_{180}|}{\sum|I_{0}|}),
\end{equation}
where $I_{0}$ and $B_{0}$ refer to the original images and the original background images of galaxies, respectively. Similarly, $I_{180}$ and $B_{180}$ refer to the $180^\circ$-rotated images and background images of galaxies. Our computation is performed within an elliptical aperture having a major axis of 1.5\,R$_{\rm p}$. The centroid of the galaxy and its neighboring pixels are used as the rotation center candidates for the computation of $A$ candidates. The smallest $A$ is taken as the rotational asymmetry of the galaxy and its corresponding rotation pixel ($x_a$,$y_a$) is the rotation center.

$C$: Light concentration, which describes the concentration of light in a two-dimensional image of a galaxy. It is usually defined as \citep[e.g.][]{Abraham1994, Bershady2000, Conselice2003a},
\begin{equation}
C=5\times \rm{log}(\frac{R_{outer}}{R_{inner}}),
\end{equation}
where $R_ {\rm outer}$ and $R_ {\rm inter}$ are the radii of the aperture, which encloses 80\% and 20\% of the total flux of a galaxy. The center of the galaxy is the rotation center ($x_a$,$y_a$). In this work, $R_ {\rm 80}$ and $R_ {\rm 20}$ are adopted to compute the $C$.

$S$: Smoothness (Clumpiness)  parameter is obtained by subtracting the smoothed image from the original image of  galaxies
\citep{Conselice2003a},
\begin{equation}
S=10 \times [(\frac{\sum(|I_{x,y}-I_{x,y}^{\sigma}|)}{\Sigma |I_{x,y}|})-(\frac{\sum(|B_{x,y}-B_{x,y}^{\sigma}|)}{\sum |I_{x,y}|})],
\end{equation}
where $I_ {x, y} $and $B_ {x, y} $ are the original image and  background image, respectively,  $I_ {x, y} ^ {\sigma} $ and $B_ {x, y} ^ {\sigma} $ indicate the smoothed image and smoothed background. The boxcar smooth kernel size is $\sigma=0.25$ $\times$ R$_{\rm p}$. The sum is carried out over all pixels at distances between 0.25\,R$_{\rm p}$ and 1.5\,R$_{\rm p}$ from rotation center \citep{Lotz2004}.

\subsubsection{$G-M_{\rm 20}$ parameters}

The $G$ parameter and the $M_{\rm 20}$ parameter are measured for pixels within the $Gini\mbox{-}segmentation$ map of unsmoothed galaxy images.

$G$: The $Gini$ coefficient is a statistical tool used in economics to quantify wealth inequality in a population. Some works have used this parameter to study the distribution of light on each pixel in galaxy images \citep{Lotz2004}. It can be computed as, 
\begin{equation}
G=\frac{1}{|\bar{f}|n(n-1)}\sum_i^n(2i-n-1)|f_{i}|,
\end{equation}
where $\bar{f}$ refers to the average flux per pixel, $n$ is the number of pixels within the $Gini\mbox{-}segmentation$ map, $i$ ranges from 0 to $n$, and $f_{i}$ is the flux of the $i$-th pixel.

A higher $G$ value indicates a more uneven distribution of light in the galaxy ($G=1$ means all the light is concentrated in a single pixel), while a lower $G$ value indicates a more uniform light distribution ($G=0$ means all pixels have same flux $\bar{f}$). \citet{Lisker2008} presents that the $Gini$ coefficient measurement is related to the image SNR. When the SNR decreases, the $G$ value approaches 0.42.

$M_{\rm 20}$: The second-order moment parameter $M_{\rm 20}$ is used to indicate the degree of spatial clustering of the brightest substructures.  The general measurement method is to arrange all pixels above a certain threshold in the galaxy image according to the flux, and then compute the normalized space second-order moment of the brightest 20\% galaxy’s flux \citet{Lotz2004} gives the formula,
\begin{equation}
M_{20}={\rm log}(\frac{\sum_i M_i}{M_{\rm tot}}),  {\rm while} \sum f_i<0.2\times f_{\rm tot},
\end{equation}
\begin{equation}
M_{\rm tot}=\sum_i^n M_i=\sum_i^n f_i[(x_i-x_m)^2+(y_i-y_m)^2],
\end{equation}
where $f_i$ is the flux value of the $i$-th pixel from the largest to the smallest within the $Gini\mbox{-}segmentation$ map.  ($x_m$, $y_m$) is the moment center pixel, which makes the $M_{\rm tot}$ minimized.

\subsubsection{$A_{\rm O}-D_{\rm O}$ parameters}

The $A$ parameter is limited in its ability to detect asymmetric structures in galaxy outskirts due to the symmetric nature of the galaxy bulge and the higher flux fraction it contains in ETGs. Additionally, extended tidal tails or faint, asymmetric structures in the galaxy outskirts are not well traced by $A$. To address these limitations, \citet{Wen2014} introduced a new non-parametric method called $A_{\rm O}-D_{\rm O}$ for detecting asymmetric structures in the galaxy outskirts. This method involves dividing the galaxy images into the outer half light region (OHR) and the inner half light region (IHR). The two parameters are obtained by computing the OHR asymmetry $A_{\rm O}$ and the relative deviation of the IHR center and OHR center $D_{\rm O}$.

First, all the pixels in the galaxy image are arranged in order of their flux, from brightest to faintest. We begin the selection of pixels from the brightest end of this arrangement, defining $f$ as the ratio of the total flux of the selected pixels to the total flux of the galaxy. As we gradually increase $f$, independent pixel groups tend to form from the brightest selected pixels. We begin by selecting pixels accounting for half of the total flux of the galaxy ($f=50\%$), and these pixels tend to form one or several independent pixel groups in images. We calculate the flux of each pixel group and continue to increase $f$ until the flux of the brightest pixel group reaches 25\% of the total flux of the galaxy. We then calculate the centroid of the brightest pixel group and use it as the center to fit an ellipse to the pixel group. We fix the axis ratio of this ellipse and gradually increase its major axis. When the flux within the elliptical aperture reaches 50\%  of the total flux of the galaxy, the ellipse is used to divide the galaxy image into IHR and OHR.

The outer asymmetry $A_{\rm O}$ is defined as follows,
\begin{equation}
A_{\rm O}=\frac{\sum{\vert I_{\rm 0} - I_{\rm 180}\vert}-\delta_2}{\sum{\vert I_{\rm 0}\vert}-\delta_1}
\end{equation}
where $\delta_1$=$f_1$*$\Sigma \vert B_{\rm 0} \vert $. $f_1$=N$_{\rm flux <1 \sigma}$/N$_{\rm all}$,  $\delta_2$=$ f_2 * \Sigma  \vert $B$_{\rm 0}$-B$_{\rm 180}\vert$, $f_2$=N$_{\vert flux\vert<\sqrt{2}}$/N$^{\prime} _{\rm all}$.  I$_ {\rm 0} $  and I$_ {\rm 180}$  refer to OHR image and 180$^{\circ}$ -rotated OHR image.  Similarly, $B_{0}$ is a background patch in the image with the same shape as $I_{0}$. The $B_{180}$ is the 180-degree rotation of the $B_{0}$. The two correction factors, $\delta_{1}$ and $\delta_{2}$, are noise contributions to the flux image $I_{0}$ and the residual image $I_{180}$, respectively. The number fraction of pixels in the OHR that are dominated by noise is represented by $f_{1}$. The $f_{2}$ represents the number fraction of OHR pixels that are dominated by noise in the residual image. The total number of pixels in the OHR and residual is represented by N$_{\rm all}$ and N$^{\prime}_{\rm all}$, respectively. The standard deviation of noise in $I_{0}$ is represented by $\sigma$.  The centroid of the whole galaxy is used as the rotational center of the OHR. See \citet{Wen2014, Wen2016} for more details.

The outer deviation $D_{\rm O}$ is defined as follows,
\begin{equation}
D_{\rm O}=\frac{\sqrt{(x_{\rm O}-x_{\rm I})^2+(y_{\rm O}-y_{\rm I})^2}}{R_{\rm eff}},
\end{equation}
where ($x_{\rm I},$ $y_{\rm I}$) and ($x_{\rm O}$,$y_{\rm O}$) refer to the centroid of the IHR and OHR, respectively. The $R_{\rm eff}$ is defined as $\sqrt{(n/\pi)}$, where $n$ is the pixel number of the IHR.

In this work, to improve the accuracy and reduce the outliers of non-parametric parameters, we first obtain two non-parametric parameter catalogs for all galaxies using the statmorph code \citep{Rodriguez-Gomez2019} and our own code, respectively. Then, we combine our derived two catalogs with the publicly available morphological parameter catalogs in the COSMOS field \citep{ Scarlata2007, Zamojski2007, Tasca2009, Wen2016} to build the best non-parametric parameter catalog by taking the median values. It should be noted that the $A_{\rm O}$ and $D_{\rm O}$ are not yet available in the publicly available morphological parameter catalogs. Thus, the values of $A_{\rm O}$ and $D_{\rm O}$ come from our measurements. In addition, the definition of $A_{\rm O}$ is slightly different between our code and the statmorph code (see \citet{Rodriguez-Gomez2019} for details).

\section{RESULTS}

\subsection{Mergers in Non-parametric Space}

\begin{figure*}[ht]
\centering
\includegraphics[width=1.0\textwidth] {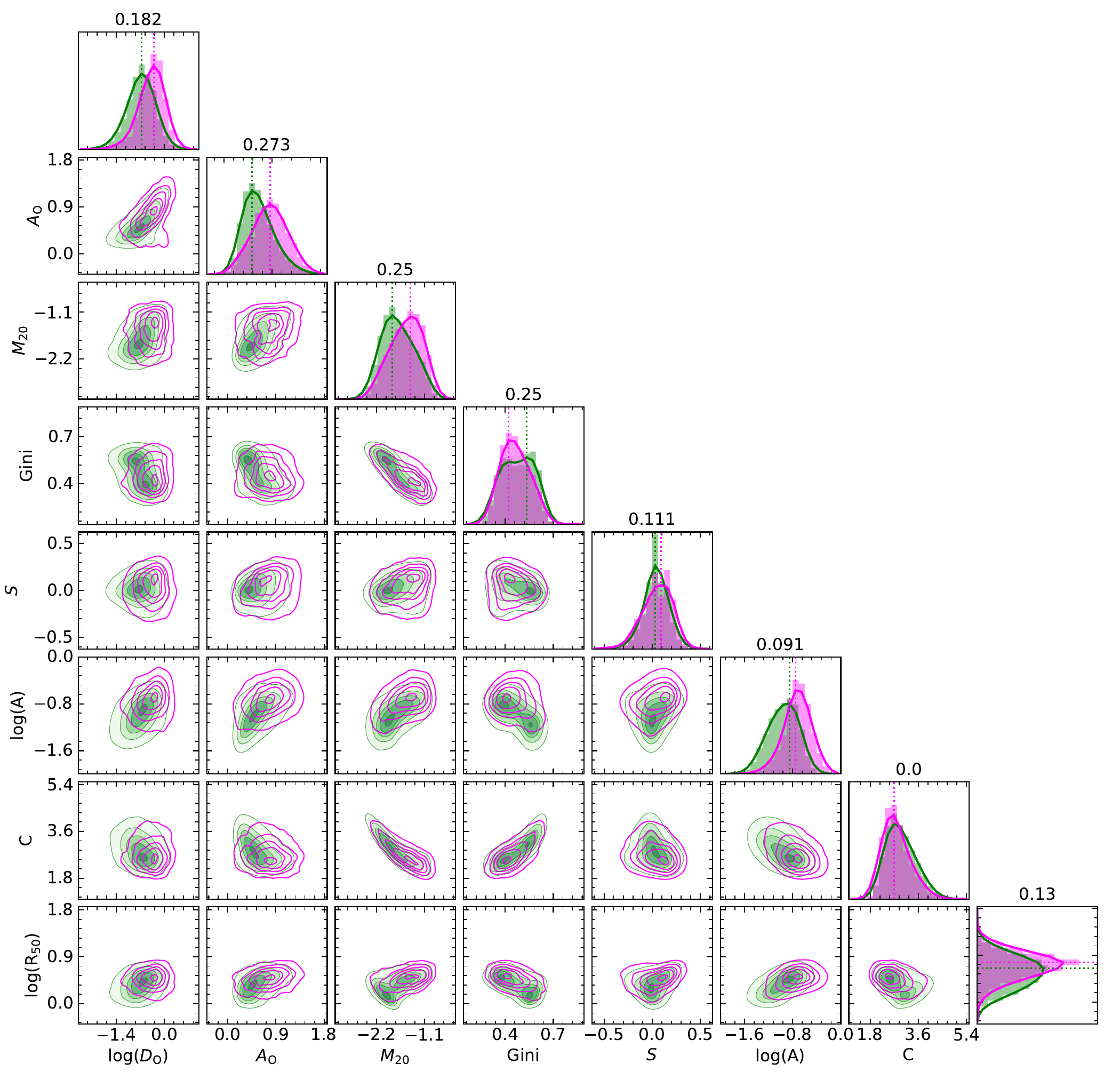} 
\caption{The distributions merger and non-merger sample in the non-parametric parameter space. The green and magenta contours refer to non-merger and merger samples. The histograms are the normalized number density of these parameters. Relative separation is given above each histogram.}
\label{fig:fig004}
\end{figure*}

The existing non-parametric parameters used for quantifying galaxy morphology can be classified into two categories. The first one measures the symmetry of light distribution in galaxies, including three parameters: $A$, $A_{\rm O}$, and shape asymmetry $A_{\rm S}$ \citep{Pawlik2016}. The second category describes the light distribution within galaxies, including parameters such as $C$, $S$, $G$, and $M_{\rm 20}$. These parameters usually trace the brighter structures within galaxies. Different parameters are used to trace the morphology of galaxies in various studies. In general, early and late-type galaxies exhibit differences in bulge size, number of star-forming clumps, and spiral arms. Therefore, parameters like $S$, $M_{\rm 20}$, and $C$ are common in the classification of early and late-type galaxies \citep{Lotz2008b}. In merging galaxies, large asymmetry is an indicator of irregular shape. Hence, the $A$ parameter plays a key role in identifying mergers. However, it should be noted that some late-type galaxies have a large number of star-forming clumps or spiral arms, which can also result in large $A$ values. Adding a $S$ parameter or $G$ parameter that is sensitive to bright substructures can help address this issue \citep{Conselice2003a}. During the late stage of mergers, there are usually two or more galactic nuclei, which can be resolved by $G-M_{\rm 20}$ \citep{Lotz2004}. Other mergers, where the tidal tails are typically fainter than the galaxy center, can be detected via the outer asymmetry parameter $A_{\rm O}$ or $A_{\rm S}$.

As merging galaxies typically exhibit morphological features different from normal Hubble-type galaxies, non-parametric parameters are often used to select merger candidates  \citep[e.g.][]{Conselice2003a, Lotz2004, Lotz2008b, Wen2014, Pawlik2016}. We removed visual mergers from the parent sample and named it the non-merger sample. Figure \ref{fig:fig004} shows the distribution of mergers and non-mergers in different parameter spaces, where the histograms represent the normalized number density distribution. The distributions of mergers and non-mergers in non-parametric parameter spaces are slightly different. To identify the most sensitive parameters for detecting the morphology of mergers, we defined a separation index $\Delta$, for example, for the $A$ parameter:
\begin{equation}
\Delta A=\frac{|A_{\rm peak}^{\rm merger}-A_{\rm peak}^{\rm non-merger}|}{|A_{95\%}^{\rm entire}-A_{5\%}^{\rm entire}|},
\end{equation}
where $A_{\rm peak}^{\rm merger}$ and $A_{\rm peak}^{\rm non-merger}$ represent the $A$ parameter values corresponding to the peaks of the normalized number density distribution of mergers and non-mergers, respectively. $A_{95\%}^{\rm entire}$ and $A_{5\%}^{\rm entire}$ represent the $A$ parameter values that contain 95\% and 5\% of galaxies in the parent sample, respectively. A larger $\Delta$ value indicates that the mergers and non-mergers are more separated in that parameter. We calculated $\Delta$ for each parameter and presented it above each histogram in Figure \ref{fig:fig004}. Our results show that $A_{\rm O}$ and $M_{20}$ are the two most sensitive parameters in identifying mergers. Although $\Delta M_{20}=\Delta G$, the distribution of the $G$ parameter has a bimodal structure in which the peaks of mergers overlap with one of the peaks of non-mergers.

After analyzing a large number of galaxy images with varying morphological parameters, we have determined that the $A_{\rm O}$ and $M_{\rm 20}$ parameters are capable of describing a wide range of galaxy morphologies. The $A_{\rm O}$ parameter is well-suited for detecting asymmetry in the galaxy outskirts and can be applied to merging galaxies, early-type galaxies, and late-type galaxies. On the other hand, the $M_{\rm 20}$ parameter is used to describe the relative distance of the brightest 20\% flux from the galaxy center. Merger events with multiple nuclei or bright star-forming regions usually have a larger $M_{\rm 20}$. Therefore, theoretically, $A_{\rm O}$ and $M_{\rm 20}$ should be able to distinguish between mergers and normal Hubble-type galaxies.

\begin{figure}[!t]
\centering
\includegraphics[width=1\columnwidth] {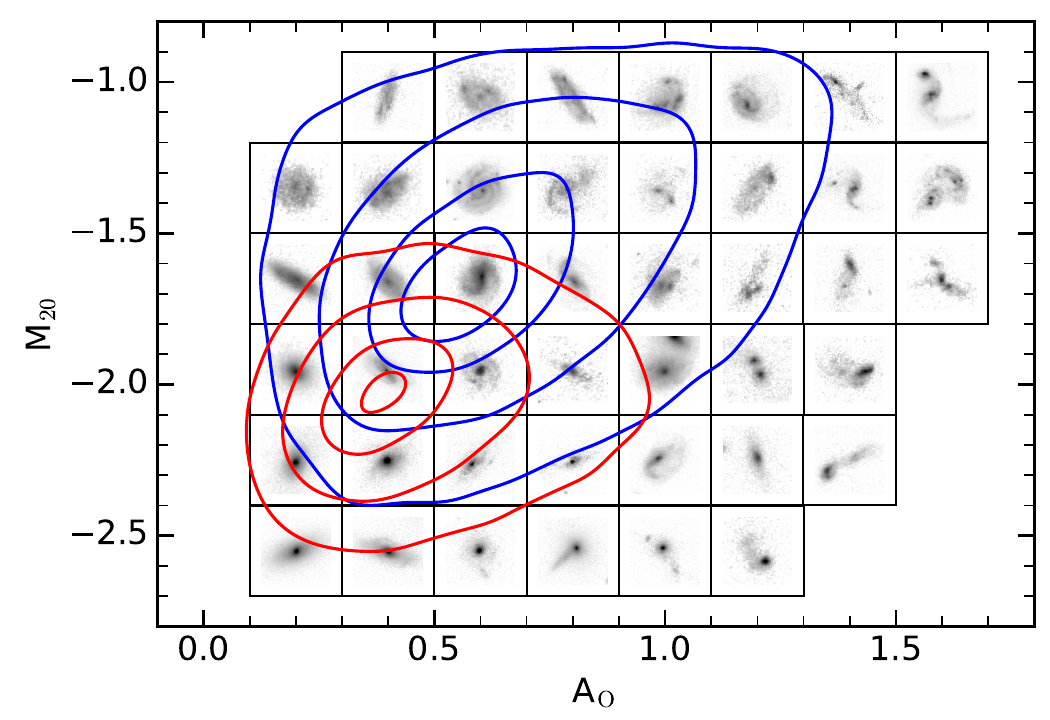} 
\caption{Example images in the $A_{\rm O}-M_{\rm 20}$ diagram. The blue (red) contours are SFGs (QGs) number density distribution.}
\label{fig:fig005}
\end{figure}

\subsection{$A_{\rm O}-M_{\rm 20}$ Merger Candidates Identification Models}

\begin{figure}[ht!]
\centering
\includegraphics[width=1\columnwidth] {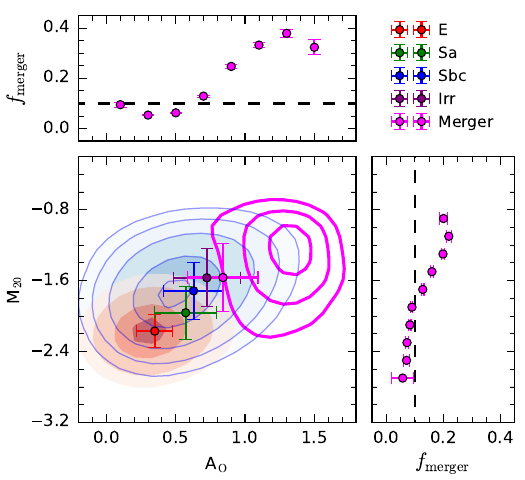} 
\caption{The distribution of SFGs, QGs, and mergers in $A_{\rm O}-M_{\rm 20}$ diagram.}
\label{fig:fig006}
\end{figure}

We visually examined galaxy images with varying $A_{\rm O}$ and $M_{\rm 20}$ values and selected the most likely morphologies to be displayed in Figure \ref{fig:fig005}. Blue and red contours in the figure refer to the number distribution of star-forming galaxies (SFGs) and quiescent galaxies (QGs) within the parent sample. Our analysis showed that images with larger $A_{\rm O}$ and $M_{\rm 20}$ parameters typically exhibit clear merger features. Figure \ref{fig:fig006} presents the median $A_{\rm O}$ and $M_{\rm 20}$ values for different Hubble-type galaxies \citep{Capak2007, Mobasher2007} and the merger fractions in the $A_{\rm O}-M_{\rm 20}$ diagram. Notably, irregular galaxies and mergers are difficult to distinguish in this diagram, as they have similar median $M_{\rm 20}$ values and very small differences in median $A_{\rm O}$ values. This similarity can lead to errors when attempting to identify merger candidates. Most of the merger candidates selected by this method are violently star-forming systems, suggesting that gas-rich wet mergers usually produce observable asymmetric tidal structures in the galaxy outskirts.

\begin{figure*}[!ht]
\centering
\includegraphics[width=1\textwidth] {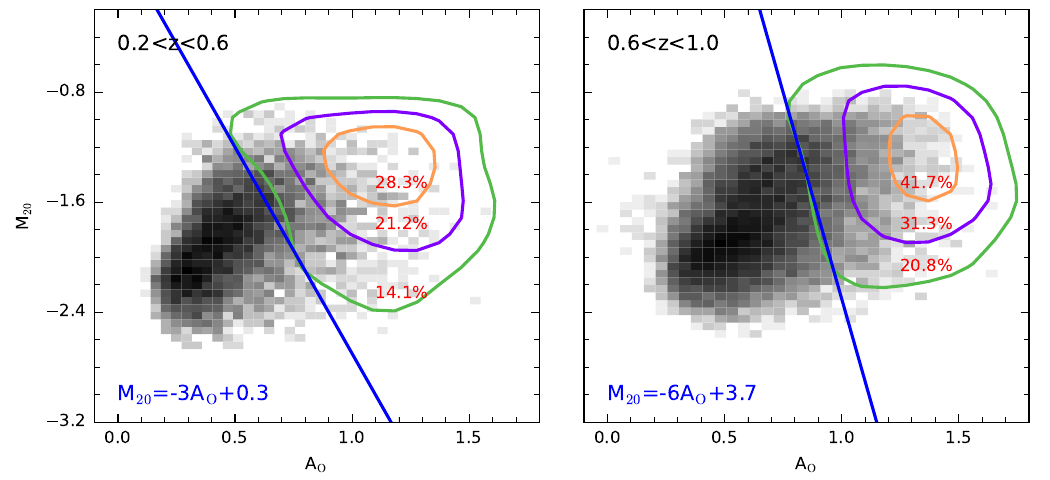} 
\caption{The distribution of merger fraction in $A_{\rm O}-M_{\rm 20}$  diagram. The black density map is the number density of the subsample. In each panel, the contours from outer to inner represent the merger fraction, which is {2, 3, 4} times the average value in each subsample masked by red numbers. The solid blue line is the merger candidate selection criterion in each subsample. The peak of the merger fraction at $A_{\rm O}=1.2$, $M_{\rm 20}=-1.3$  in low-$z$ subsample and $A_{\rm O}=1.4$,  $M_{\rm 20}=-1.3$  in high-$z$ subsample.}
\label{fig:fig007}
\end{figure*}

To reduce the dependence of the merger fraction and non-parametric morphological parameters on redshift, we divided the parent sample and merger sample into low-$z$ ($0.2< z<0.6$) and high-$z$ ($0.6<z<1.0$) subsamples. We then split the $A_{\rm O}-M_{\rm 20}$ diagram into $30\times 30$ grids and calculated the merger fraction for each grid within each subsample, obtaining two merger fraction density maps. We subsequently smoothed these maps by applying a two-dimensional Gaussian kernel with $\sigma_{M_{\rm20}}=0.22$ and $\sigma_{A_{\rm O}}=0.152$, which is related to $A_{\rm O}$ and $M_{\rm 20}$ errors observed in different wavelengths. More details can be found in the Appendix.
Figure \ref{fig:fig007} presents our results, which indicate that the merger fraction in each subsample increases with higher $A_{\rm O}$ and $M_{\rm 20}$ values, but the number density of galaxies reduces. Using lines where the merger fraction is more than twice the average value enables us to distinguish normal galaxies from merger candidates in each subsample. About 44\% of visual samples can be classified as mergers in our non-parametric diagnostics. These mergers are violent gas-rich mergers showing strong tidal features.

\subsection{Merger Rate}

We have only 19.5\% of spec-$z$+spec-$z$ pairs in our pair sample. Some true pairs are excluded in our sample due to the scatter of phot-$z$ in the samples selected using phot-$z$. The corrected pair number can be expressed as
\begin{equation}
N_{\rm pair}=N_{\rm s+s}+Q_1 N_{\rm s+p}+Q_2 N_{\rm p+p}.
\end{equation}

The $Q_1$ and $Q_2$ are obtained by Monte Carlo methods as follows: \\
1. We calculated the fraction of phot-$z$ in our sample at each redshift, $f(z) = N_{\rm phot-z}(z)/N_{\rm total}(z)$.  \\
2. We assumed that $z_{\rm best}$ in our sample is the true redshift, without any errors. \\
3. We randomly added to the redshifts of the galaxies in the above sample with a Gaussian-distributed random error $\Delta z$, with $\sigma = 0.005(1 + z)$, to be taken as the simulated photometric redshift. The redshifts of the remaining galaxies are taken as spec-$z$. \\
4. We used the method in Table \ref{tab: pair samples} to obtain the simulated spec-$z$ + phot-$z$ galaxy pair $N_{\rm p+s}$ and phot-$z$ + phot-$z$ galaxy pair $N_{\rm p+p}$.
Then, we obtain the number of true pairs $N'_{\rm p+s}$  and $N'_{\rm p+p}$  identified by $|\Delta V|<500$ km\,s$^{-1}$ without $\Delta z$ in the $N_{\rm p+s}$ and $N_{\rm p+p}$ samples.
We define $Q_1= N'_{\rm p+s}$/$N_{\rm p+s}$ and $Q_2= N'_{\rm p+p}$/$N_{\rm p+p}$.\\
5. Steps 3 and 4 were repeated 1000 times. The  $Q_1$ and $Q_2$ are obtained by taking the median results of the 1000 simulations. \\

Using the Monte Carlo method, we estimate the value of $Q_1$ and $Q_2$ to be approximately 1.5. Since our pair sample is complete for galaxies with log(M$_*$/M$_\odot)>10.1$, we describe the merger fraction calculated using the pair sample as:
\begin{equation}
f_{\rm merg,pair}=\frac{N_{\rm pair}}{N_{\rm galaxy}(\log(M_*/M_\odot)>10.1)}.
\end{equation}

\begin{figure*}[]
\centering
\includegraphics[width=1.0\textwidth] {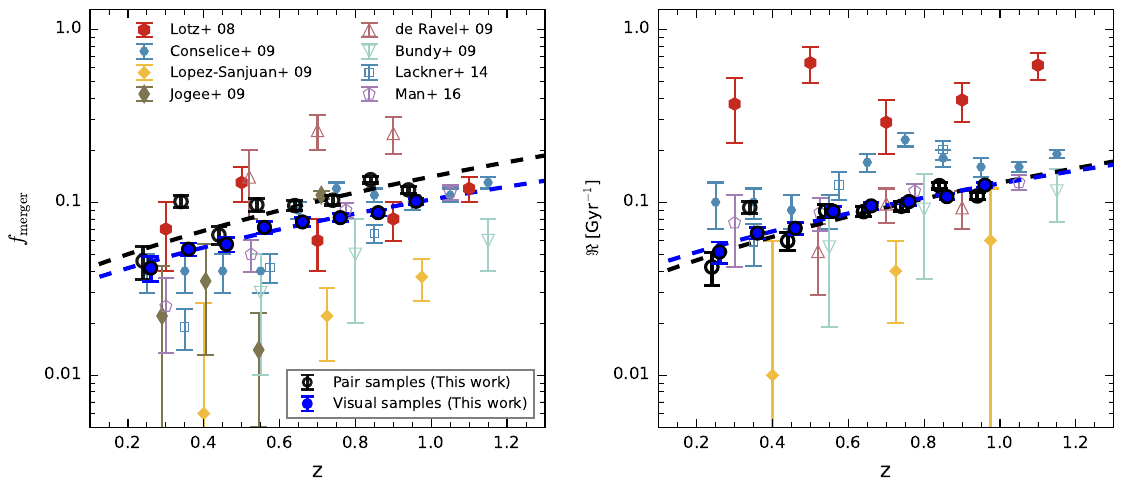} 
\caption{ Merger fraction and merger rate as a function of redshift. The filled markers represent major mergers selected by morphological methods \citep{Lotz2008a, Conselice2009, Lopez-Sanjuan2009, Jogee2009}, whereas the unfilled markers denote massive close pairs \citep{deRavel2009, Bundy2009, Lackner2014, Man2016}. The blue-filled and black-unfilled markers refer to the visual and pair samples in our work, respectively. The blue and black dashed lines represent the best-fit relation of our visual and pair samples, respectively. The error bars of our markers are statistical errors.}
\label{fig:fig008}
\end{figure*}

To examine the evolution of the merger fraction with redshift, we divided the galaxy sample into eight bins and used our merger and pair samples to determine the merger fraction for each bin. Additionally, we gathered available data on the mass-limited merger fraction evolution for $z<1.2$, employing morphological and galaxy pair methods. The left panel of Figure \ref{fig:fig008} displays the redshift evolution of the merger fraction for $z<1.2$. Unfilled markers indicate pair samples, while filled markers represent samples identified via the morphological method (visual and non-parametric).  We then fit the evolution of $f_{\rm merger}$ with redshift, $f=(0.035\pm 0.008) (1+z)^{2.02\pm 0.42}$ for the pair sample, and $f=(0.032\pm 0.002) (1+z)^{1.79\pm 0.13}$ for the visual merger sample, respectively. These results are consistent with the latest studies conducted on intermediate redshifts \citep{Thibert2021}.

The galaxy merger rate can be expressed as the number of mergers occurring per unit co-moving volume per time. However, different number densities of galaxies in distinct sky regions, caused by cosmic variance, can create uncertainties when determining the merger rate $\Gamma$. Consequently, it is challenging to compare the evolution of merger rates across various studies. To address this, we employed the fractional merger rate $\Re = f_{\rm merger} / \mathrm{T}_{\rm obs}$ to evaluate the merger rate's evolution. Here, $f_{\rm merger}$ denotes the merger fraction in each redshift bin. T$_{\rm obs}$ is an observable merger timescale, which varies among different merger samples and is typically estimated from numerical merger simulations. Several studies such as \citet{Jogee2009}, \citet{Bridge2010}, \citet{Wen2016}, and \citet{Whitney2021} proposed merger timescales of roughly $0.5-0.8$\,Gyr for visually identified mergers. Our visual merger samples lasted for a long time from the early pair stages to the final merger stages. Therefore, we calculated $\Re$ for the visual merger sample by adopting T$_{\rm obs}=0.7$\,Gyr.

The merger rate based on galaxy pairs is defined by \citet{Man2016},

\begin{equation}
\Re_{\rm merg, pair}=\frac{C_{\rm merg,pair}\times f_{\rm pairs}}{T_{\rm obs,pair}},
\end{equation}
where, $C_{\rm merg, pair}$ is a multiplicative factor that indicates the fraction of pairs likely to merge within the $T_{\rm obs,pair}$ interval \citep{Lotz2011}. For our $R_{\rm proj}=5-50$\,kpc selected galaxy pairs, we employed $C_{\rm merg,pair}$=0.6 and $T_{\rm obs,pair}$=0.65 Gyr to compute the merger rate \citep{Lotz2010a, Mantha2018}. Further information on $T_{\rm obs,pair}$ and $C_{\rm merg,pair}$ can be found in previous studies \citep[e.g.][]{Kitzbichler2008, Man2016, Lotz2010a, Lotz2010b, Lotz2011, Mantha2018}.

In the right panel of Figure \ref{fig:fig008}, we plot our visual and pair sample merger rates. The markers in the panel are the same as those in the left panel. Our results suggest that there is little difference between the visual merger sample and the pair sample in each redshift bin, indicating that our findings are independent of the merger galaxy sample selection method. Consequently, we obtain a more realistic evolutionary trend of merger rates with redshift for $z<1$. We fit the data to determine the evolutionary relation $f=(0.032\pm 0.007) (1+z)^{2.02\pm 0.42}$ and $f=(0.047\pm 0.003) (1+z)^{1.79\pm 0.13}$ for the pair sample and the visual merger sample, respectively. For galaxies with log($M_*$/M$_\odot) > 9.5$ and $z < 1$, we estimate that they experience $\sim$0.1 major mergers per Gyr.

\subsection{Star Formation and AGN Activity in Mergers}
We utilize the merger sample and pair sample to investigate the impacts of mergers on both the star formation and AGN activity of galaxies. Firstly, we separate star-forming galaxies (SFGs) from quiescent galaxies (QGs) for the parent sample using the rest-frame $UVJ$ diagram \citep{Williams2009},
\begin{equation}
\begin{split}
(U-V)>0.88 \times (V-J)+0.69 \quad [0.0<z<0.5],  \\
(U-V)>0.88 \times (V-J)+0.59 \quad [0.5<z<1.0] .
\end{split}
\end{equation}
Additional criteria of $(U-V)>1.3$ and $(U-V)< 1.6$ are applied in the two redshift bins.
Then, we cross-matched the parent sample with a publicly available AGN catalog \citep{Delvecchio2017} to obtain a total of 431 AGNs. Finally, we divided the parent sample into three sub-samples: the merger sample (3594 galaxies), the pair sample (3474 galaxies), and the non-interaction sample (26775 galaxies).

\subsubsection{Star formation properties}

\begin{figure*}[]
\includegraphics[width=1.0\textwidth] {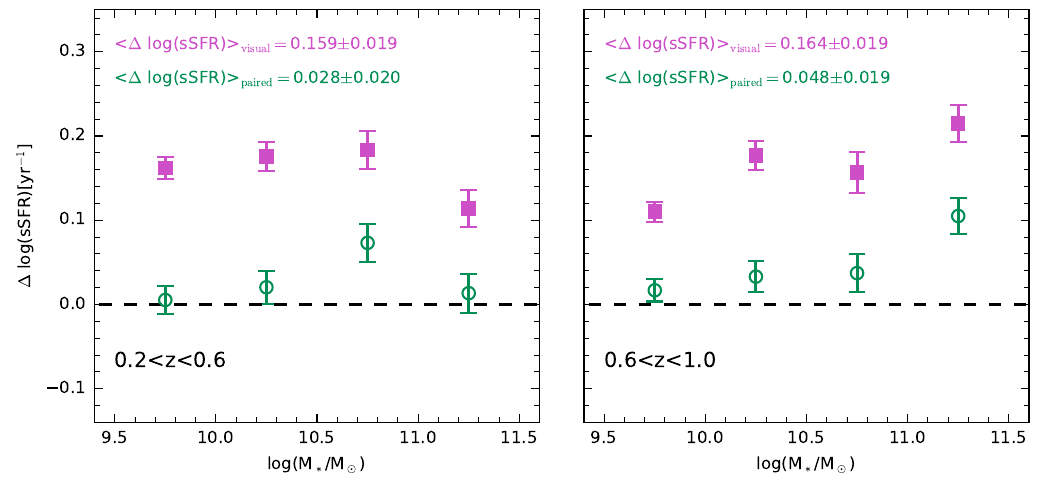} 
\caption{Star formation rate excess as a function of stellar mass in mergers. Magenta and green markers indicate the visual and pair samples, respectively. The left panel represents the low-$z$ subsample, while the right panel represents the high-$z$ subsample.}
\label{fig:fig009}
\end{figure*}

Observations and simulations are still inconclusive about whether mergers drive star formation activity in intermediate to high redshift galaxies. Some studies have shown that the impact of galaxy mergers on star formation activity varies across different redshifts. To investigate this, we divide the visual sample, pair sample, and non-interaction sample into low-$z$ and high-$z$ subsamples and study the star formation properties of SFGs in each subsample.

We employed the specific star formation rate (sSFR) to trace the star formation activity of galaxies. Each subsample is divided into four mass bins. The star formation excess is calculated using the following equation,
\begin{equation}
\Delta\log(\rm{sSFR})=\log(\rm{sSFR})_{\rm non\mbox{-}interaction}-\log(\rm{sSFR})_{\rm merger}.
\end{equation}
We first randomly pick up a merger from the visual merger sample with a given mass and redshift bin, then randomly pick up a galaxy from the non-interaction sample in the same mass and redshift bin to calculate the $\Delta$log( sSFR). This process was repeated 1000 times to obtain the median $\Delta$log( sSFR). It should be noted that the median $\Delta$log( sSFR) represents the separation of the merger and non-interaction samples on the star-forming main sequence (SFMS), while the error represents the dispersion of the SFMS. We repeat these steps 1000 times to obtain the mean $<$$\Delta$log( sSFR)$>$ and its error. We utilize the same method to estimate the $<$$\Delta$log( sSFR)$>$ and error for the pair sample.

Our results are presented in Figure \ref{fig:fig009}. Magenta and green points indicate the sSFR excess in the visual and pair samples relative to non-interaction galaxies, respectively. Points above the black dashed line suggest an enhancement of star formation activity in the merging galaxy, while points below the line are indicative of mergers that inhibit star formation. We find that the visual merger sample demonstrated an overall higher specific star formation rate than the non-interaction sample by approximately 0.16\,dex. In contrast, the pair sample showed almost no excess, primarily because we selected the visual merger sample are gas-rich mergers. By contrast, the pair sample is observed during the early stage of mergers when the enhancement effect was not noticeable.  Furthermore, we should note that our pair sample is complete for pairs with log($M_*$/M$_\odot)>$10.1. Galaxies smaller than this mass are part of companion galaxies for massive galaxies. Thus, satellite quenching effects might play a dominant role in star formation activity for the smaller galaxies in pairs.

\subsubsection{AGN enhancement in mergers}

We will investigate the effect of galaxy mergers on AGN activity by comparing the AGN fraction in the visual merger sample, pair sample, and non-interaction sample. We compute the AGN fraction in different stellar mass and redshift bins for all samples. Then, we calculate the AGN fraction excess in the visual merger sample and pair sample using the following formula,
\begin{equation}
\Delta f_{\rm AGN}=f_{\rm AGN}^{\rm merger}-f_{\rm AGN}^{\rm non\mbox{-}interaction}.
\end{equation}

\begin{figure}[]
\centering
\includegraphics[width=1\columnwidth] {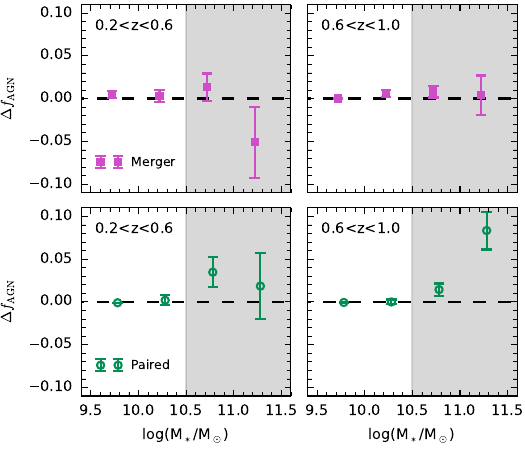} 
\caption{AGN fractions excess in different stellar mass bins for the visual merger sample and the pair sample. The shadow region in each panel refers to the massive bins (log($M_*$/M$_\odot$)$>10.5$).}
\label{fig:fig010}
\end{figure}

In Figure \ref{fig:fig010}, magenta and green markers represent the visual merger samples and pair samples, respectively. Our results show no significant AGN excess in the visual merger sample which suggests that mergers do not trigger AGN activity in the visual merger sample. We propose that this is mainly because the visual merger sample is dominated by mergers of late-type galaxies, where the central black hole is less massive. Hence it triggers less AGN activity during the merger process. 

Meanwhile, massive galaxies (log($M_\ast$/M$_\odot$)$>10.5$) in low redshift have a relatively high AGN fraction \citep{Kauffmann2003}, which makes the AGN fraction in the visual merger sample lower than that in the non-merger sample at the massive low-$z$ bin. However, the AGN fractions in high-mass bins of the pair sample are relatively high, with 1-3 $\sigma$ higher than that of the non-interaction sample in massive bins. This suggests that interaction may trigger AGN activity in massive galaxies in pair-method selected mergers.

\section{Discussions} \label{sec:Discussion}

\subsection{The Merger Sample}   

Different merger selection methods can result in samples with different characteristics. For example, \citet{Lackner2014} selected 2055 late-type galaxies from the COSMOS field, whereas \citet{Wen2016} used a combination of non-parametric and visual methods to identify 461 merging galaxies with long tidal tails. These features are harder to form or last a shorter period, resulting in a smaller sample size. In contrast, our merger sample includes remnants from pairs of galaxies with tidal features until the end of the merger, resulting in a longer observable timescale and a larger sample size. Although merger sample sizes vary between studies, the observed merger rate should be divided by the observed timescale to ensure a consistent evolution of the merger rate with redshifts.

It should be noted that F814W corresponds to the rest-frame U-band at relatively high redshifts. For star-forming galaxies, dust extinction is more severe, and more clumps tend to form, leading to the misidentification of such galaxies as mergers. Although the COSMOS-DASH survey \citep{Mowla2019} obtained F160W images of these galaxies, the depth and resolution were lower than those of F814W images, making identification challenging. However, the high-resolution and high-sensitivity COSMOS-Web images in the near-infrared could help obtain more precise merger samples in the future \citep{Casey2022}.

Only 19.5\% of our pairs have spectroscopic redshifts. In the case of photometric redshift pairs, large errors may occur in high-density environments. Nevertheless, our results mostly agree with those of other studies.

\subsection{Parameter Measurements}
\label{Parameter Measurements}

The measured morphological parameters of galaxies in F814W images are subject to systematic bias and diffuseness at different redshifts for three main reasons. Firstly, the morphology of F814W images differs in different rest-frame wavelengths. For galaxies at $0.2<z<0.6$, the rest-frame wavelengths correspond to $678-509$\,nm. At $0.6<z<1$,  the rest-frame wavelengths are at $509-407$\,nm. As our sample changes from the rest-frame V-band at low redshift to the rest-frame U-band at high redshift, the morphology of the same galaxy may vary considerably between bands. Secondly, the morphology and structure of galaxies vary at different redshifts due to their gas content, different Hubble-type galaxy fractions, star formation, dust extinction, and intrinsic evolution \citep{Baes2020, Yao2023}. All of these factors lead to changes in the morphological parameters. Thus, it is necessary to study the evolution of the morphological parameters with redshift at the same rest-frame wavelength.  Finally, the measurement of morphological parameters is similarly affected by the cosmic dimming effect. This effect not only affects non-parameter methods but also other methods used to detect faint structures. Two effective methods to address this issue include increasing the exposure time or investigating the effect of different surface brightness limits on parameter measurements from simulated images.

The best way to use non-parametric parameters is within the same survey \citep{Holwerda2021}. The image SNR, the FWHM of the PSF, the cosmic dimming effect, and the pixel scale all have an important effect on $A_{\rm O}$ as well as other parameters. In general, for the same galaxy, the larger the FWHM of the PSF, the smaller the measured $A_{\rm O}$. The lower the image’s SNR, the larger the scatters in the measured $A_{\rm O}$ value. For the redshift effect, \citet{Yu2023} investigated the redshift and PSF effects by redshifting the DESI observed galaxies to {\it JWST} CEERS observations.  Actually, the best way to evaluate the effects of SNR and PSF on the morphological parameters is to use realistic mock galaxies by adding different noises and different FWHMs of PSF. We will investigate the corrections for the observational effects on HST and JWST images using simulated galaxies in another forthcoming work soon.

Besides, performing ``morphological k-correction" is a key step when studying samples with a wide redshift range. For $A$ parameters, 
\begin{equation}
A=A_{\rm obs}+\delta A_{\rm SB-dim}+\delta A_{\rm k-corr}+A_{\rm noise}+...
\end{equation}
\citet{Lopez-Sanjuan2009} find that the asymmetry parameter increases with redshift with $\delta A=-0.05z$ for $M_{\rm B}<-20$ galaxies. Using multi-wavelength data from {\it HST}, \citet{Conselice2008} find $\delta A_{\rm k-corr}<0.2$ for galaxies with $z<1.2$. \citet{Baes2020} showed that non-parametric parameters vary significantly from the $UV$ to submillimeter bands for nearby galaxies. Future investigations using multi-wavelength imaging data from {\it JWST} will help us correct the variability of single-band morphological parameters by studying the effect of different rest-frame bands on the non-parametric parameters.

We divide our sample into low-$z$ and high-$z$ subsamples to reduce the influence of wavelength on the merger classification results. We give different merger selection criteria in different redshift bins. Our criteria are the rest-frame V-band selection condition in the low-$z$ bin and the rest-frame U-band selection condition in the high-$z$ bin. We realize that it makes sense to combine the selection criteria at different redshifts, and the consequence of doing so would be to involve redshifts in the selection criteria. We hope that more high-resolution multi-wavelength imaging data from subsequent COSMOS fields will address this issue.

\subsection{The Non-parameter Merger Selection Model}

\begin{figure*}[]
\centering
\includegraphics[width=1\textwidth] {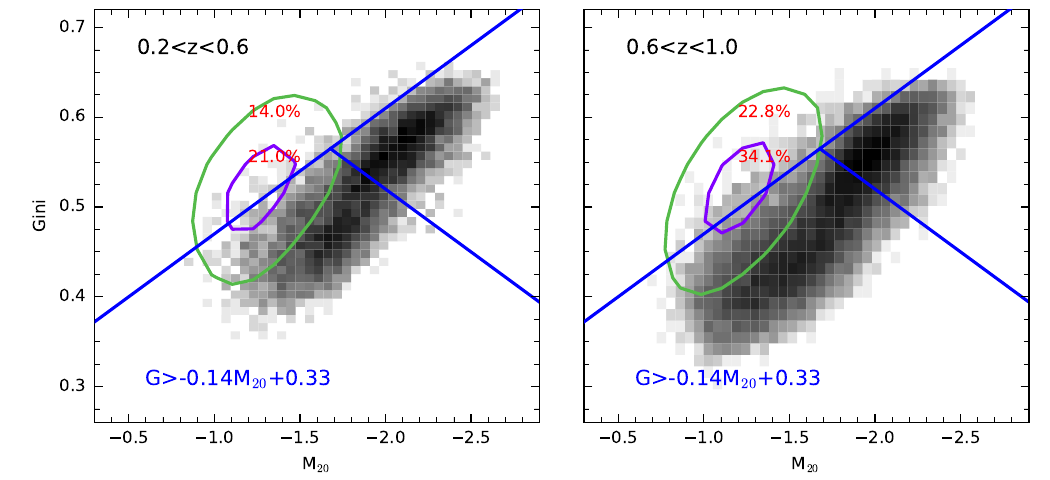} 
\caption{The merger fraction in $Gini-M_{\rm 20}$ diagram. Contours from outer to inner represent the merger fraction, which is higher than 2,3 times the mean value in the parent sample. }
\label{fig:fig011}
\end{figure*}  

Previous works use Several non-parametric morphological methods to identify merging galaxy candidates. We tested the widely used $Gini-M_{\rm 20}$ method on our merger sample. To do this, we calculated the distribution of the merger fraction on the $Gini-M_{\rm 20}$ diagram using the same method as Figure \ref{fig:fig007}. Figure \ref{fig:fig011} shows the results, with the blue line indicating the classification criteria provided by \citet{Lotz2011} for mergers (top left), early-type galaxies (top right), and late-type galaxies (bottom right). The contours represent the distribution of the merger fraction.
The mergers we identified were mainly gas-rich, so some of them fell in the region where the late-type galaxies of the $Gini-M_{\rm 20}$ diagram are located. This type of galaxy is not well identified by the $Gini-M_{\rm 20}$ method. In contrast, our $A_{\rm O}-M_{\rm 20}$ method provides a criterion based on all visual mergers and is better suited for identifying gas-rich mergers. This is primarily because the $A_{\rm O}$ parameter is more sensitive to tidal features in gas-rich mergers.

Non-parametric merger galaxy identification methods with one or two parameters are only able to select a small fraction of mergers. A multi-parameter approach is necessary to significantly improve the accuracy and completeness of merging galaxy identification. As shown in Figure \ref{fig:fig004}, the ability to distinguish merging galaxies from normal galaxies varies with different parameters, which suggests that using machine learning to give a method for selecting mergers weighted by each parameter would be effective. For instance, \citet{Snyder2015a} used S($G$, $M_{\rm 20}$) $=0.139 M_{\rm 20} + 0.990 G-0.327$ to analyze mergers.

Similarly, in the $A_{\rm O}$, $M_{\rm 20}$ parameter space, the merger fraction is high in regions with high $A_{\rm O}$ and $M_{\rm 20}$.  We suspect that at higher redshifts, $A_{\rm O}$ may be a better parameter for identifying merging galaxies than other parameters. This is mainly because high redshift galaxies have a high fraction of gas-rich galaxies, which are more likely to produce observable tidal structures in the galaxy outskirts. Based on our results, galaxies with $A_{\rm O}>0.9$ have shown signatures of mergers. Hence, $A_{\rm O}=0.9$ could be considered the boundary between mergers and non-mergers.

\subsection{The Role of Merger in Galaxy Evolution}

Many Ultra-Luminous Infrared Galaxies (ULIRGs) exhibit merger features and AGN in the nearby Universe \citep{Sanders1988, Urrutia2008}. Studies of galaxy pairs in the low redshift Universe have found that mergers can enhance star formation compared to isolated galaxies \citep{Patton2013, Darg2010, Ellison2013, Barrera-Ballesteros2015}. However, this effect is significantly weaker in the intermediate to high redshift Universe \citep{Perret2014, Fensch2017, Patton2020}. In a recent study, \citet{Shah2022} uses an interacting sample containing 2351 spectroscopic redshift pairs to investigate the enhancement of star formation in the CANDELS fields at $0.5<z<3$.  They find that galaxy interaction does not significantly enhance star formation. Besides, \citet{Pearson2019} uses the ML method to identify mergers and non-mergers from more than 200,000 galaxies and find that the SFR enhancement factor of $\sim 1.2$ in the merger sample at $z<4$. Our work shows that visual mergers enhance the specific star formation rate (sSFR) by only about 0.16\,dex (a factor of roughly 1.4) compared to the non-interaction sample. Although our visual merger sample is independent of the sample selected by the two aforementioned studies, both yield relatively consistent results. This suggests that at intermediate redshifts, major mergers indeed have little impact on star formation in galaxies. The results indicate a negligible contribution of major mergers to the cosmic star formation rate density (CSFD) at $z<1$.

\section{conclusions}
 \label{sec:Summary}

In this work, we construct a volume-limited stellar mass complete galaxy sample using the latest photometric and spectroscopic redshift data from the COSMOS field. This sample contains 33065 galaxies with log(M$_\ast $/M$_\odot )\geq 9.5$ and $0.2\leq z \leq 1.0$. Among them, 13881 galaxies have spectroscopic redshifts. The photometric redshift accuracy is $\sigma_{\rm NMAD} \sim0.005$. We measure the non-parametric morphological parameters of {\it HST}/ACS F814W images for sample galaxies and analyze the properties of merging galaxies in this sample. Our main results are summarized as follows:

\begin{itemize}

\item [\textbf{1.}]  We use the visual method and pair method to identify a visual merger sample containing 3594 galaxies and a pair sample containing 1737 massive galaxy pairs. The merger rates as a function of redshifts with a slope of 2.02$\pm$ 0.42 for the pair sample and 1.79$\pm$0.13 for the visual merger sample at $z<1$.

\item [\textbf{2.}] By analyzing the non-parametric morphological parameters of mergers and non-merger galaxies, we find that the outer asymmetry parameter $A_{\rm O}$ and the second order moment parameter $M_{\rm 20}$ are the two most effective parameters to identify star-forming mergers. We propose the selection criteria to identify the candidates of violently star-forming merging galaxies based on $A_{\rm O}$ and $M_{\rm 20}$. The identification formula are $M_{\rm 20}>-3A_{\rm O}+0.3$ for $0.2<z<0.6$ galaxies and $M_{\rm 20}>-6A_{\rm O}+3.7$ for $0.6<z<1.0$ galaxies.  

\item [\textbf{3.}] We find that the sSFR of the visual merger sample is higher than that of the non-interaction sample by $\sim0.16$\,dex (a factor of $\sim$ 1.4). There is a weak enhancement of star formation activity caused by major mergers. However, there is no star formation enhancement in paired galaxies.  This means that the effects of galaxy mergers on star formation differ at different stages of merging. The enhancement in the visual merger sample has a negligible contribution to the CSFD at $z<1$. The AGN fraction in the visual merger sample is consistent with that in the non-interaction sample at all redshift and mass bins. The AGN fraction in massive paired galaxies is 1-3 $\sigma$ higher than that of non-merger. It is likely that visually selected merging galaxies contain more star-forming galaxies, where the black holes are less massive and less likely to trigger AGN during the merging process.
\end{itemize}

This work boasts higher completeness compared to previous methods for selecting merging galaxies and massive close pairs. We exploited more accurate redshifts to correct for contamination from foreground and background galaxies. In addition, we developed a new non-parametric method to effectively identify the candidates of violently star-forming mergers. Our visual merger sample can be used as a training set to develop machine-learning models.

Furthermore, we intend to use the $A_{\rm O}$ and $M_{\rm 20}$ parameters, which are sensitive not only to the morphology of merging galaxies but also to different regions of the galaxy, to investigate the connections between galaxy morphologies and their other properties. In the near future, we will use these parameters to bridge the morphological evolution of galaxies from intermediate to high redshifts, as observed by $JWST$, $CSST$, and $Roman$ space telescope.
Our non-parametric morphological catalogs in the COSMOS field will be released publicly, after completing the non-parametric measurements of COSMOS-Web near-infrared images. Before that, the current non-parametric products in {\it HST}/ACS F814W are available upon request to the authors. 

\section{acknowledgements}

The authors thank the anonymous referee for the useful and detailed comments and suggestions. This project is supported by the National Natural Science Foundation of China (NSFC grants No.12273052,11733006), the science research grants from the China Manned Space Project (No. CMS-CSST-2021-A04). NL acknowledges the support from the Ministry of Science and Technology of China (No. 2020SKA0110100), the science research grants from the China Manned Space Project (No. CMS-CSST-2021-A01) and the CAS Project for Young Scientists in Basic Research (No. YSBR-062). XXZ acknowledges the support from the National Science Foundation of China (12233005 and 12073078), the science research grants from the China Manned Space Project ( No. CMS-CSST-2021-A02, CMS-CSST-2021-A04 and CMS-CSST-2021-A07). Some of the data presented in this paper were obtained from the Mikulski Archive for Space Telescopes (MAST) at the Space Telescope Science Institute. The specific observations analyzed can be accessed via\dataset[DOI:10.17909/T9XW2Q]{https://doi.org/DOI:10.17909/T9XW2Q}. This research made use of \texttt{Astropy}, a community-developed core Python package for Astronomy \citep{AstropyCollaboration2013}.

\bibliography{References_merger}{}
\bibliographystyle{aasjournal}

\begin{figure*}[]
\centering
\includegraphics[width=1\textwidth] {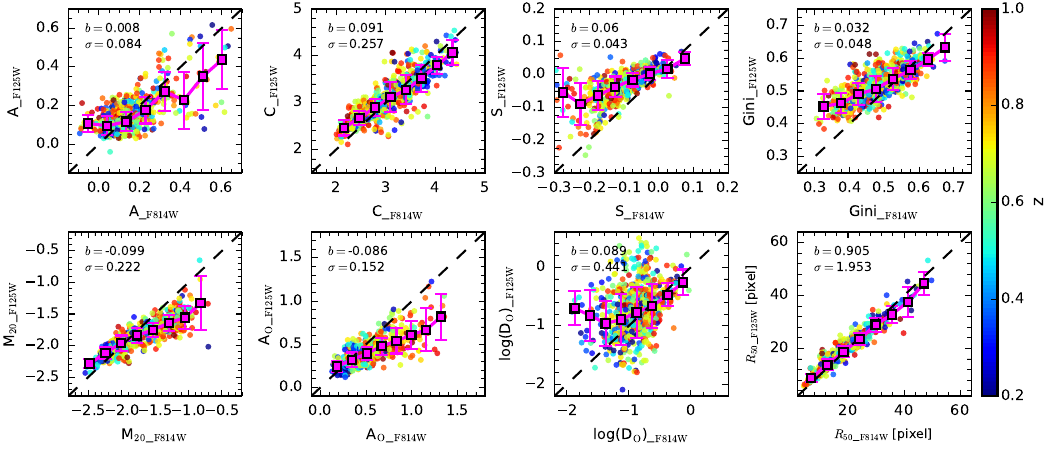} 
\caption{A comparison of the non-parametric morphological parameters in the F814W and the F125W images.}
\label{fig:appendix}
\end{figure*}

\appendix
\label{appendix}
We match our sample with the CANDELS F125W imaging data \citep{Koekemoer2011} and obtain a sample of 1191 galaxies, which have F814W images and F125W images. Then the F814W parameters and F125W parameters of these galaxies were calculated using our code. Since the rest-frame band of F125W at $0.6<z<1$ just corresponds to the rest-frame band of F814W at $0.2<z<0.6$.  Our results are shown in Figure \ref {fig:appendix}. It is clear that there is a large dispersion in the parameters calculated for different bands, and for some of the band-sensitive parameters, there is a large deviation in the relatively high redshift. For $M_{\rm 20}$, which we used, the overall deviation is small despite some dispersion. The $A_{\rm O}$ is much more sensitive to the wavelength. This is why we give different criteria conditions for merging galaxies in different redshift bins. 

\end{document}